\documentclass[%
 reprint,
showpacs,
 amsmath,amssymb,
 aps,
 pra,
 showkeys,
longbibliography
]{revtex4-1}

\usepackage{graphicx} % Include figure files
\usepackage{dcolumn} % Align table columns on decimal point
\usepackage{bm} % bold math

\usepackage[utf8]{inputenc}
\usepackage[T1]{fontenc}
\usepackage{times}
\usepackage{flafter} % Place floats after their first reference
\usepackage{textgreek} % use non-italic lowercase greek characters
\usepackage{mathrsfs} % mathscr for fancy floquet H
\usepackage[usenames,dvipsnames]{xcolor}
\usepackage[hidelinks,urlcolor=blue, colorlinks=true,citecolor=blue, linkcolor=blue ]{hyperref}

\newcommand{\ii}{ i }
\newcommand{\ee}{\mathrm{e}}

\renewcommand{\Re}{\operatorname{Re}}

\newcommand{\fullwf}{\mathit{\Phi}}
\newcommand{\velel}{{\hat v_\mathrm{ee}}}

\newcommand{\pr}{^\prime}

\newcommand{\odd}{\text{ odd}}
\newcommand{\even}{\text{ even}}

\newcommand{\Vee}{\hat{v}_\mathrm{ee}}
\newcommand{\Ge}{\hat{\mathit{\Gamma}}_\mathrm{e}}
\newcommand{\tr}{\mathrm{Tr}}
\newcommand{\ket}[1]{| #1 \rangle}
\newcommand{\bra}[1]{\langle #1 |}
\newcommand{\ketbra}[2]{| \vphantom{#2} #1 \rangle\langle \vphantom{#1} #2 |}

\newcommand{\com}[2]{\left[#1 \vphantom{#2} , #2 \vphantom{#1}  \right]}
\newcommand{\acom}[2]{\left[#1 \vphantom{#2} , #2 \vphantom{#1}  \right]_{+}}

\newcommand{\braket}[2]{\langle #1 \vphantom{#2} |  #2 \vphantom{#1} \rangle} % for Dirac brackets
\newcommand{\fbraket}[3]{\langle #1 \vphantom{#2#3} | #2 | #3 \vphantom{#1#2} \rangle}
\newcommand{\abs}[1]{|#1|}

\newcommand{\iek}[1]{\left(#1\right)}
\newcommand{\kiek}[1]{\left[#1\right]}
\newcommand{\fiek}[1]{\left\{#1\right\}}

\newcommand{\ketrno}[1]{|\tilde{#1} \rangle}

\newcommand{\mat}[1]{\mathbf{\mathcal{#1}}}

\makeindex

\begin{document}

%\title{Fano profiles and high-harmonics in time-dependent renormalized-natural-orbital theory}
\title{Single-photon double ionization: renormalized-natural-orbital theory {\em vs} multi-configurational Hartree-Fock}
%\thanks{A footnote to the article title}%

\author{M.\ Brics}
\author{J.\ Rapp}
\author{D.\ Bauer}%
\affiliation{%
 Institut für Physik, Universität Rostock, 18051 Rostock, Germany
}

\date{\today}

\begin{abstract}
The $N$-particle wavefunction has too many dimensions for a direct time propagation of a many-body system according to the time-dependent Schrödinger equation (TDSE). On the other hand, time-dependent density functional theory (TDDFT) tells us that the single-particle density is, in principle, sufficient. However, a practicable equation of motion (EOM) for the accurate time evolution of the single-particle density is unknown. It is thus an obvious idea to propagate a quantity which is not as reduced as the single-particle density but less dimensional than the $N$-body wavefunction. Recently, we have introduced time-dependent renormalized-natural-orbital theory (TDRNOT). TDRNOT is based on the propagation of the eigenfunctions of the one-body reduced density matrix (1-RDM), the so-called natural orbitals. In this paper we demonstrate how TDRNOT is related to the multi-configurational time-dependent Hartree-Fock (MCTDHF) approach. We also compare  the performance of MCTDHF and TDRNOT {\em vs} the TDSE  for single-photon double ionization (SPDI) of a 1D helium model atom. SPDI is one of the effects where TDDFT does not work in practice, especially if one is interested in correlated photoelectron spectra, for which no explicit density functional is known.
\end{abstract}


\maketitle

\section{Introduction}
\label{sec:intro}
Double ionization by electron or photon impact  is a prime example for a correlated atomic process \cite{Schmidt-Bocking2003,Berakdar2003,Ciapp10,amusia2013atomic}.
In single-photon double ionization (SPDI), one photon is absorbed by a multi-electron system, followed by the emission of two electrons. As the laser-atom interaction term in the Hamiltonian involves only one-electron operators, the photon interacts only with one electron directly, whose energy can then be shared via Coulomb interaction with another electron. SPDI of helium has been studied for 50 years now \cite{ByronJoachain} and is still of interest to date \cite{Forre12}. Thanks to the increasing availability of free electron laser (FEL) sources, time-resolved studies of correlated or collective processes following the absorption of an XUV photon are within reach now (see, e.g., \cite{Lemell15}).

In this work, we employ SPDI as a demanding benchmark for the recently developed  time-dependent renormalized-natural-orbital theory (TDRNOT) \cite{tdrnot, tdrnot2, tdrnot3, tdrnot4, tdrnot5} and the widely known  multi-configurational time-dependent Hartree-Fock (MCTDHF) \cite{Zanghellini2004, Scrinzi_mctdhf, Hochstuhl2010}. We work out the connection between TDRNOT and MCTDHF and benchmark their performance with a 1D helium model atom, for which the time-dependent Schrödinger equation (TDSE) is still numerically exactly solvable.

The development of powerful time-dependent quantum many-body approaches beyond linear response is one of the great challenges in theoretical and computational physics. In fact, a full numerical solution of the time-dependent Schrödinger equation (TDSE) for strong-field problems in full dimensionality is impossible for more than two electrons \cite{Scrinzi2016}. Unfortunately, efficient methods such as time-dependent density functional theory (TDDFT) \cite{UllrichBook, Ullrich2013} fail, in particular if strong correlations are involved \cite{florian2, rabi-1}. In TDDFT, also some of the observables of interest cannot be expressed explicitly in terms of the single-particle density \cite{florian,florian2}.  It is thus an obvious idea to propagate a quantity which is not as reduced as the single-particle density but less dimensional than the wavefunction.  Prominent candidates for such quantities are reduced density matrices \cite{Coleman2000,AppelPhDThesis,GiesbertzPhDThesis,Lackner2015,Lackner2017}.

In recent years, we introduced a novel method that goes one step beyond TDDFT as far as the complexity of the propagated quantity is concerned. In TDRNOT \cite{tdrnot, tdrnot2, tdrnot3, tdrnot4, tdrnot5}, the basic quantities that are propagated are the eigenfunctions of the one-body reduced density matrix (1-RDM), normalized to their eigenvalues, the so-called  renormalized natural orbitals (RNO). There are also other wavefunction-based approaches available in the literature which overcome the problems of TDDFT \cite{Ishikawa2015,Hochstuhl2014}. The most frequently used are MCTDHF \cite{Zanghellini2004, Scrinzi_mctdhf, Hochstuhl2010} and  time-dependent configuration interaction (TDCI) \cite{Greenman2010, Pabst2013, Karamatskou2014}.

The paper is organized as follows.  The theoretical method and the connection between TDRNOT and  MCTDHF is described in Sec.~\ref{sec:theor}. In Sec.~\ref{sec:model}, we introduce the 1D helium model atom which is used as a benchmark system in Sec.~\ref{sec:res} to compare the performance of TDRNOT and MCTDHF regarding SPDI. We conclude in Sec.~\ref{sec:concl}.

Atomic units (a.u.) are used throughout unless noted otherwise.

\section{Theoretical methods}
\label{sec:theor}
In this Section, we relate the recently introduced TDRNOT to MCTDHF by deriving the equations of motion (EOM) for both.

The time evolution of the $N$-particle state $\ket{\fullwf (t)}$ is described by the TDSE
\begin{equation}
\ii \ket{\dot\fullwf (t)} =\hat H(t)\ket{\fullwf (t)}.
\label{eq:TDSE}
\end{equation}
The EOM for the $N$-body density matrix ($N$-DM) of a pure state
\begin{align}
  \hat\gamma_N(t)
    &=
\ketbra{\fullwf(t)}{\fullwf(t)} \label{eq:dm}
\end{align}
is obtained by taking the time derivative of \eqref{eq:dm} and using the TDSE \eqref{eq:TDSE}.

With an $N$-particle Hamiltonian of the form
\begin{equation}
 \hat H(t)=\sum_{i=1}^N \iek{\hat h^{(i)}(t)-\ii\Ge^{(i)}} +\sum_{i<j}^N \hat v_{\mathrm{ee}}^{(i,j)}
\end{equation}
where $\hat h(t)$ is the hermitian part of the single-particle Hamiltonian consisting of kinetic energy, electron-nucleus interaction, and electron interaction with external fields, e.g., the laser field, $-\ii\Ge$ is an imaginary potential for absorbing outgoing electron flux, and $\hat v_{\mathrm{ee}}^{(i,j)}$ is the electron-electron interaction potential where the upper indices indicate that the operator is acting on electrons $i$ and $j$, the EOM for the $N$-DM reads
\begin{equation}
 \begin{split}
   \ii \dot{\hat\gamma}_N(t) =& \sum_{i=1}^N\com{\hat h^{(i)}(t)}{\hat \gamma_N(t)}+\sum_{i<j}^N \com{\Vee^{(i,j)}}{\hat \gamma_N(t)}\\
   &-\ii\sum_{j=1}^N \acom{\Ge^{(j)}}{\hat \gamma_N(t)}
 \end{split}
 \label{eq:eomndm}
\end{equation}
where $[\hat a,\hat b]$ and $[\hat a,\hat b]_+$ are commutator and anti-commutator of two operators $\hat a$ and $\hat b$, respectively. By applying partial traces of \eqref{eq:eomndm} one can derive EOMs for the $n$-RDMs
\begin{equation}
\hat \gamma_n(t) = \binom {N} {n} \mathrm{Tr}_{n+1,\dots,N} \hat\gamma_N(t).
\end{equation}
The EOM for 1-RDM reads
\begin{equation}
\begin{split}
 \ii \dot {\hat \gamma}_1(t) =& \com{\hat h(t)}{\hat\gamma_{1}(t)}
  + 2\tr_{2} \com{\Vee}{\hat\gamma_{2}(t)} -\ii \acom{\Ge}{\hat\gamma_{1}(t)}\\ &-2\ii
\tr_{2} \acom{\Ge^{(2)}}{ \hat \gamma_{2}(t)}.
\label{eq:eom_g1}
\end{split}
\end{equation}
The EOM \eqref{eq:eom_g1} requires the knowledge of the 2-RDM. Similarly the EOM for the 2-RDM involves the 3-RDM and so on. The resulting system of coupled equations is known as the  BBGKY hierarchy (Bogoliubov, Born, Green, Kirkwood, Yvon) \cite{Bogoliubov1946, Bogoliubov1947, Yvon1935, Kirkwood1946, Kirkwood1947, Born1946} and is more complicated to solve than the TDSE. Thus any application in practice aims at truncating the hierarchy at some level $q < N$. As in our case $N=2$ and we do not want to propagate the 2-RDM (which is of twice the number of dimensions of the $2$-electron wavefunction) we cut the BBGKY hierarchy already after the first equation \eqref{eq:eom_g1}. However, $\hat\gamma_1(t)$ still has the same dimensionality as the  $2$-electron wavefunction. We therefore  expand $\hat\gamma_1(t)$ and $\hat\gamma_2(t)$ in a complete, orthonormal basis of single-particle orbitals $\hat 1=\sum_{n=1}^\infty\ketbra{n(t)}{n(t)}$, $\braket{m(t)}{n(t)}=\delta_{mn}$,
\begin{align}
 \hat \gamma_1(t)&=\sum_{mn}\rho_{mn}(t)\ketbra{m(t)}{n(t)} \label{eq:gamma1exp},\\
\hat \gamma_2(t)&=\sum_{ijkl}\gamma_{2,ijkl}(t)\ketbra{i(t)j(t)}{k(t)l(t)} \label{eq:gamma2exp},
\end{align}
where the shorthand notation for tensor products $\ket{i(t)j(t)}=\ket{i(t)}^{(1)}\ket{j(t)}^{(2)}=\ket{i(t)}^{(1)}\otimes\ket{j(t)}^{(2)}$ is used, and a superscript index indicates the particle to which states refer.
Note that the expansion coefficients are connected via
\begin{equation}
 \rho_{mn}(t) =\frac{2}{N-1} \sum_j \gamma_{2,mjnj}(t)
\end{equation}
and  are formally defined as
\begin{align}
 \rho_{mn}(t)&=\fbraket{m(t)}{\hat \gamma_{1}(t)}{n(t)} \label{eq:rhomn},\\
 \gamma_{2,ijkl}(t)&=\fbraket{i(t)j(t)}{\hat \gamma_{2}(t)}{k(t)l(t)} \label{eq:gamma2ijkl}.
\end{align}
By inserting the expansions $\eqref{eq:gamma1exp}$ and \eqref{eq:gamma2exp} into \eqref{eq:eom_g1} the EOM for the time-dependent orbitals is obtained, which turns out to be the same as the EOM for MCTDHF orbitals,
\begin{equation}
\begin{split}
\ii\ket{\dot n (t)}=&\hat R(t)\Biggl[\iek{\hat h(t)-\ii\Ge} \ket{n(t)} \\&+ 2\sum_{ijkl}\rho_{kn}^{-1}(t)   \gamma_{2, ijkl}(t) \fbraket{ l(t)}{\hat v_{\mathrm{ee}}}{j}\ket{ i(t)}\Biggr] \\&+\sum_{j} g^T_{nj}(t)\ket{j(t)},
\label{eq:MCTDHF}
\end{split}
\end{equation}
where $g_{mn}(t)=\fbraket{m(t)}{\hat g (t)}{n(t)}=\ii\braket{m(t)}{\dot n(t)}$  with an arbitrary hermitian operator $\hat g(t)$. The sums in \eqref{eq:MCTDHF} are finite now and run over the $N_\circ$ orbitals considered in the numerical implementation, which span a truncated subspace.  The operator  $\hat R(t)= \hat 1 - \sum_{n=1}^{N_\circ} \ketbra{n(t)}{n(t)}$  projects onto the orthogonal complement of that subspace.

 In order to solve \eqref{eq:MCTDHF} numerically an expression for $\gamma_{2,ijkl}(t)$  and a convention for $\hat g(t)$ need to be chosen.
Regarding $\gamma_{2,ijkl}(t)$, one approach is to expand the state in the same truncated orthonormal basis as  $ \hat \gamma_1(t)$ and  $ \hat \gamma_2(t)$,
\begin{equation}
\ket{\fullwf(t)}=\sum_{j_1\dots j_N}d_{j_1\dots j_N }(t)\ket{j_1(t)\cdots j_N(t)},
\label{eq:wfexpan}
\end{equation}
and propagate the $\binom{N_\circ}{N}$ non-zero and independent coefficients $d_{j_1\dots j_N }(t)$, which are formally defined as
\begin{equation}
 d_{j_1\dots j_N }(t)=\braket{j_1(t)\cdots j_N(t)}{\fullwf(t)}.
\end{equation}
The EOM for the expansion coefficients $d_{j_1\dots j_N }(t)$ is  obtained by inserting \eqref{eq:wfexpan} into the TDSE \eqref{eq:TDSE} and multiplying from the left with $\bra{j_1(t)\cdots j_N(t)}$.
% \begin{equation}
% \ket{\fullwf(t)}=\sum_{j_1j_2\dots j_N}d_{j_1j_2\dots j_N }(t)\ket{j_1(t)}\ket{j_2(t)}\cdots\ket{j_N(t)}
% \end{equation}
Then $\gamma_{2,ijkl}(t)$ can be expressed as the partial trace
\begin{equation}
\gamma_{2,ijkl}(t)=\binom {N}{n} \sum_{m_3\dots m_N}d_{ijm_3\dots m_N }(t)d^*_{klm_3\dots m_N }(t).
\end{equation}
However, propagating the objects $d_{j_1\dots j_N }(t)$ with $N$ indices, each running over the number of orbitals taken into account, seems unnecessary expensive considering that all the information needed for propagation is contained in the 4-index object $\gamma_{2,ijkl}(t)$. It would be desirable to write $\gamma_{2,ijkl}(t)$ in terms of an even less dimensional quantity with known EOM. In the special case of two particles, there exists an exact and adiabatic mapping from $\rho_{mn}(t)$ to $\gamma_{2,ijkl}(t)$ which is used~\cite{tdrnot2} in TDRNOT for $N = 2$ and given below as~\eqref{eq:gamma2ijkl_}. For $N > 2$, useful approximations to $\gamma_{2,ijkl}(t)$ are the ``holy grail'' of natural-orbital theory. Candidates to be tested are, e.g., PNOF5e~\cite{PNOF5e} and PNOF6($N_c$)~\cite{PNOF6}.

Regarding $\hat g(t)$, the particular gauge choice $\hat g^{\mathrm{NO}}(t)$ defined in the next paragraph relates the MCTDHF EOM to the TDRNOT EOM as long as the exact expression for $\gamma_{2,ijkl}(t)$ is retained. The role of $\hat g(t)$ has already been described in Refs.~\cite{Meyer1990, Manthe1992} in the context of the multi-configurational time-dependent Hartree approach, including a debate whether the particular choice of $\hat g^{\mathrm{NO}}(t)$ is beneficial or not~\cite{Jansen1993, Manthe1994}. In principal, any choice $\hat g(t)$ should give the same result. In practice, the simulation may benefit from a gauge choice leading to EOMs with better numerical properties; for instance, small matrix elements  $g_{mn}(t)$ might allow for larger time steps. Common gauge conventions are $g_{ij}(t)=0$ or $g_{ij}(t)=\fbraket{j(t)}{\hat h(t)}{i(t)}$, where $g_{ij}(t)=0$ usually allows to use slightly larger time steps.

The particular $\hat g^{\mathrm{NO}}(t)$ is defined such that the orbitals $|n(t)\rangle$ are eigenfunctions  of the 1-RDM, called natural orbitals (NOs), i.e.,
\begin{equation}
\hat \gamma_1(t)=\sum_k n_k(t) \ketbra{k(t)}{k(t)} \label{gamma1inNOs}
\end{equation}
and $\rho_{mn}(t)=\delta_{mn}n_n(t)$, where $n_k(t)$ are the corresponding eigenvalues,  called occupation numbers (ONs). This is possible because the matrix elements $\rho_{mn}(t)$ depend on the gauge choice,
\begin{equation}
\begin{split}
\ii\dot\rho_{mn}(t)=&\ii\fbraket{m(t)}{\dot\gamma_1(t)}{n(t)}-\sum_{k}\rho_{kn}(t)g_{mk}(t)\\ &+\sum_{k}\rho_{mk}(t)g_{kn}(t),
\end{split}
\end{equation}
which is obtained by taking the time derivative of \eqref{eq:rhomn} and inserting unities $\hat 1 =\sum_k\ketbra{k(t)}{k(t)}$.

For $n\neq m$ if $n_n(t)\neq n_m(t)$, one finds out that for NOs
\begin{equation}
 g^{\mathrm{NO}}_{mn}(t)=\frac{\ii\fbraket{m(t)}{\dot\gamma_1(t)}{n(t)}}{n_n(t)-n_m(t)}.
\end{equation}
Note that when  $n_n(t)= n_m(t)$ all terms $g^{\mathrm{NO}}_{mn}(t)$ are undetermined. In this case eigenvalues are degenerate and any orthogonal pair of eigenstates from the subspace they span can be selected. For those terms any value generated by some arbitrary hermitian operator $\hat g(t)$ can be chosen (we use $g^{\mathrm{NO}}_{mn}(t)=0$). Also, all diagonal terms $g^{\mathrm{NO}}_{mm}(t)$ are undetermined because the phases of the NOs (as eigenstates of the 1-RDM) are not defined. Here we use  the phase convention presented in \cite{tdrnot4},
\begin{equation}
\begin{split}
 \ii \braket{n}{\dot n}=&\frac{1}{2}\fbraket{n}{\hat h(t)}{n}+\frac{1}{2}\fbraket{n'}{\hat h(t)}{n'} \\&+  \frac{1}{n_n(t)}\Re \sum_{jpl} \gamma_{2,plnj}(t) \fbraket{ nj  }{ \Vee}{pl} ,
\end{split}
 \end{equation}
which ensures that for two-electron systems $\gamma_{2,ijkl}(t)$ is an adiabatic functional of the ONs \cite{tdrnot2},
\begin{equation}
\begin{split}
  \gamma_{2, ijkl}(t)&={d}_{ij}(t){d}^*_{kl}(t)\delta_{i,j'}\delta_{k,l'} \\&=\iek{-1}^{i-k}\sqrt{n_i(t)n_k(t)}\frac{\ee^{\ii\kiek{\varphi_i-\varphi_k}}}{2}\delta_{i,j'}\delta_{k,l'},
  \label{eq:gamma2ijkl_}
  \end{split}
\end{equation}
where the ``prime operator'' acts on a positive integer $k$ as
\begin{align}
  k\pr
    &=
      \begin{cases}
        k+1 & \text{if $k\odd$}\\
        k-1 & \text{if $k\even$},
      \end{cases}
  &
  k
    &>
      0,\label{eq:prime-operator}
\end{align}
and $\ee^{\ii\varphi_i}$ are phase factors which, if one allows for complex groundstate  NOs, can be set to $\ee^{\ii\varphi_i}=1$.

If one chooses to propagate NOs,  one can propagate either the set of NOs and the expansion coefficients for the wavefunction $d_{j_1\dots j_N }(t)$  (as in MCTDHF) or the set of NOs and ONs, using the exact expression or an approximation for  $\gamma_{2,ijkl}(t)$.  For two-electron systems, it turns out that the second choice is numerically more efficient. Moreover, the propagation according to the EOM for the combined quantity %
\begin{equation}
\ket{\tilde k(t)}=\sqrt{n_k(t)} \ket{k(t)},
\end{equation}
called renormalized  NOs (RNOs),
is more stable.
The EOM for the RNOs  read  \cite{tdrnot4}
\begin{equation}
\begin{split}
  \ii|\dot{\tilde n}\rangle
    =&
      \iek{\hat{h}(t)-\ii\Ge}\ketrno{n}+ {\mat{A}}_n(t) \ketrno{n}\\
       &+ \sum_{k\neq n} {\mat{B}}_{nk}(t)\ketrno{k}
       + \sum_k {\mat{\hat{C}}}_{nk} (t) \ketrno{k}, \label{eq:eom-rno}
\end{split}
\end{equation}
with
\begin{equation}
\begin{split}
  \mat{{A}}_n(t)
    =&
      -\frac{1}{{n}_n(t)}\Re\sum_{jkl}
      \tilde\gamma_{2,njkl}(t)
      \langle \tilde k \tilde l |\velel|\tilde n \tilde j \rangle \\
      &-\frac{1}{2{n}_n(t)}\iek{\fbraket{\tilde n}{\hat{h}(t)}{\tilde n}-\fbraket{\tilde n'}{\hat{h}(t)}{\tilde n'}}\\
&-2\ii  \sum_{jl} \tilde
\gamma_{2,njnl}(t) \fbraket{\tilde l}{\Ge}{\tilde j} ,
\end{split}
\end{equation}
\begin{align}
 {\mat{\hat{C}}}_{nk}(t)
    &=
      2\sum_{jl} \tilde \gamma_{2,kjnl}(t)\langle \tilde l|\velel| \tilde j \rangle, \label{eq:eom-end}
\end{align}
\begin{align}
\begin{split}
 &{\mat{B}}_{nk}(t)
    =
     \frac{\fbraket{\tilde k(t)}{ \sum_{p}{\mat{\hat{C}}}_{np}(t)}{\tilde p(t)}-\fbraket{\tilde  n(t)}{ \sum_{p}{\mat{\hat{C}}}_{kp}(t)}{\tilde p(t)}^*}{{n}_n(t)-{n}_k(t)} \\
        & -4\ii\frac{n_n(t)}{n_n(t)-n_k(t)}\sum_{jl}\tilde \gamma_{2, kjnl}(t) \fbraket{\tilde l}{\Ge}{\tilde{j}}  \\
& -2\ii\frac{1}{n_n(t)-n_k(t)}\fbraket{\tilde k}{\Ge}{\tilde n} , \quad n_k(t) \neq n_n(t),
\label{eq:Bnk1}
\end{split}
\end{align}
%
%
%and
%
%
and
\begin{equation}
\tilde \gamma_{2,ijkl}(t)=\frac{1}{\sqrt{n_i(t)n_j(t)n_k(t)n_l(t)}} \gamma_{2,ijkl}(t).
\end{equation}
In summary, there are three essential steps from the general EOM~\eqref{eq:MCTDHF} to the EOM for RNOs being propagated in TDRNOT. First, a functional for $\gamma_{2,ijkl}(t)$ is used, which for $N=2$ is known exactly but for $N>2$ needs to be approximated.  Second, $\hat g^{\mathrm{NO}}(t)$ is chosen to make the orbitals $|n(t)\rangle$ natural. Finally, the NOs are renormalized to their occupation number, yielding the RNOs $|\tilde n(t)\rangle = \sqrt{n_n(t)}\,|n(t)\rangle$.

Our numerical investigations show that it is very important to use the EOM \eqref{eq:eom-rno} with the imaginary potential taken properly into account. For example,  we observed in \cite{tdrnot2} that  during Rabi oscillations the NO  with the lowest ON among all NOs taken into account in the numerical propagation shows erratic behavior after a while, subsequently spoiling NOs with higher ONs.
 In \cite{tdrnot2} we thought this effect is due to the necessary  truncation of the number of NOs considered during propagation, due to which the last NO cannot couple correctly to all other NOs. Now we know that with the EOM~\eqref{eq:eom-rno} properly accounting for the antihermitian part $-\ii\Ge$ in the Hamiltonian to absorb outgoing electron flux, no erratic behavior occurs. These findings should be also relevant if a mask function instead of an imaginary potential is used \cite{Sato2014}. Alternatively, infinite-range exterior complex scaling \cite{Scrinzi2000} could be used if high absorption efficiency over small grid distances is required.

\section{Model  atom}
\label{sec:model}
We employ the  widely used  one-dimensional helium model atom  \cite{Eberly, Haan1994, Dieter97, Lappas_Leeuwen,Lein2005} for benchmarking. The Hamiltonian reads
\begin{align}
  \hat H^{(1, 2)}(t)
    &=
      \hat h^{(1)}(t)
      + \hat h^{(2)}(t)
      + \velel^{(1, 2)}-\ii\Ge^{(1)}-\ii\Ge^{(2)} \label{hamiltonian}
\end{align}
where upper indices indicate the action on either electron $\mathrm 1$, electron $2$, or both. The single-particle Hamiltonian in dipole approximation and velocity gauge (with the purely time-dependent $A^2(t)$ term transformed away) reads
\begin{align}
  \hat h(t) &=
      \frac{\hat p^2}{2}
      - \frac{2}{\sqrt{\hat x^2 + \varepsilon_{\mathrm{ne}}}}+A(t)\hat p,
\end{align}
and the electron-electron interaction is given by
\begin{align}
  \velel^{(1, 2)}
    &=
      \frac{1}{\sqrt{\left(\hat x^{(1)}-\hat x^{(2)}\right)^2 + \varepsilon_\mathrm{ee}}}.
\end{align}
For the imaginary potential we chose
\begin{equation}
\Ge=50\, (\hat x / x_{\mathrm{B}})^{16}
\end{equation}
where $\mp x_\mathrm{B}$ denote the coordinates of the left and right boundaries of the 1D grid, respectively. All calculations were performed for the spin-singlet configuration, starting from the ground state.
The values for the parameters $\varepsilon_\mathrm{ne}=0.50$ and $\varepsilon_\mathrm{ee}=0.33$ were thus chosen to match the real, three-dimensional He and He$^+$ ionization potentials. Because of the separability of the wavefunction into spin and spatial components, the number of spatial RNOs that actually need to be propagated reduces to $N_\circ^\mathrm{spat}=N_\circ/2$.

\section{Results}
\label{sec:res}
Results from  TDRNOT and MCTDHF calculations for SPDI, together with the corresponding TDSE benchmark, will be presented in this Section. All results were obtained starting from the spin-singlet ground state, which was calculated via imaginary-time propagation. Finite differences on an equidistant real-space grid with $1024$ grid points (in each spatial direction) and a grid spacing of $0.2$ have been employed. An adaptive time step via the Dormand–Prince RK 5(4) method \cite{rk54dp} was used in the MCTDHF and TDRNOT calculations.

\subsection{Single-photon double ionization }
\label{sec:spdi}
\begin{figure}[htbp]
%\includegraphics[width=0.99\columnwidth]{p2dim_tdse_proj_}
%\hfill
\includegraphics[width=0.99\columnwidth]{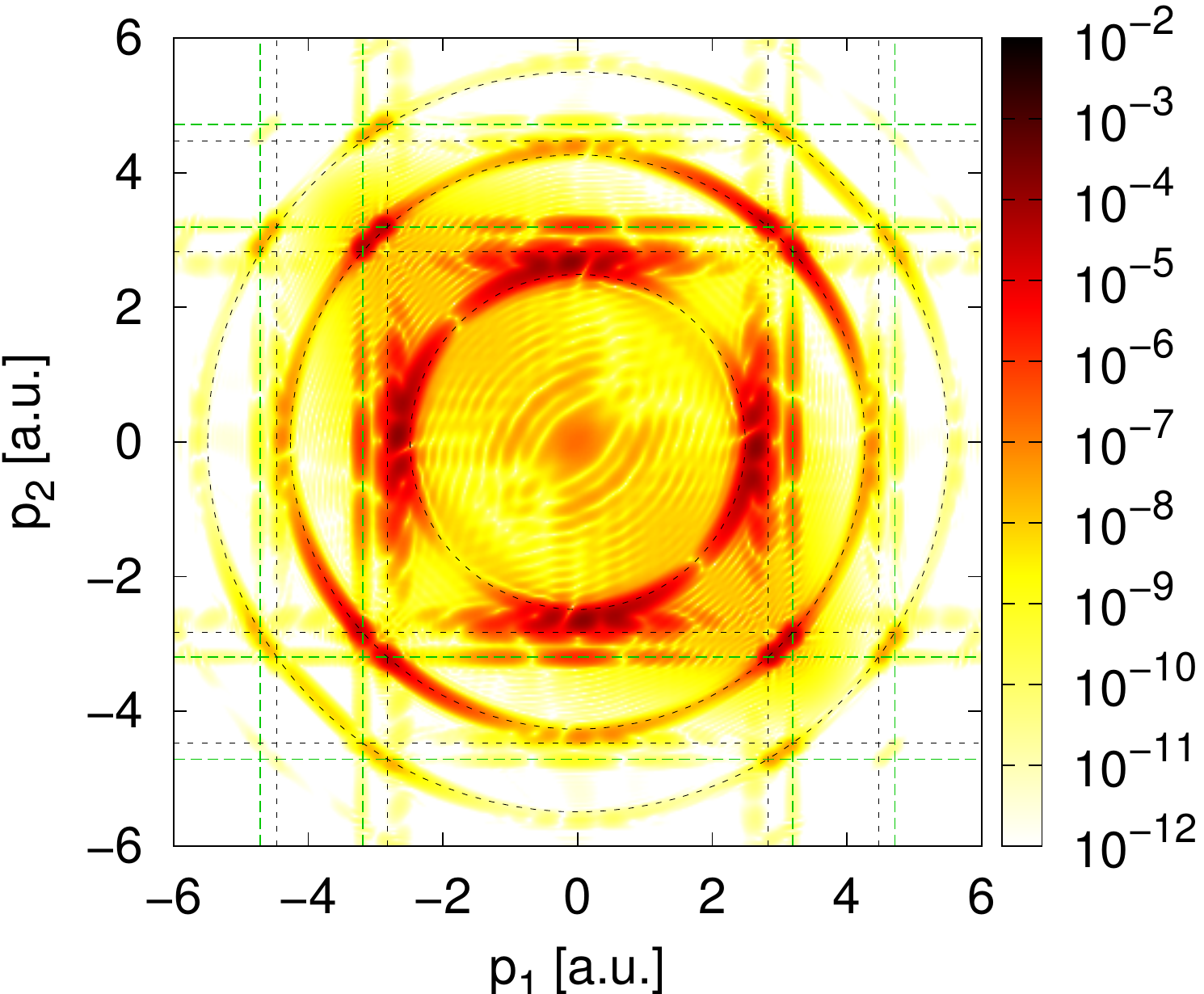}
\caption{(Color online) Correlated photoelectron momentum spectrum calculated from the TDSE by applying a filter in position-space \cite{florian2, tdrnot3}. A 7.6-nm 20-cycle $\sin^2$-shaped laser pulse of peak intensity  $I_0=3.2 \times 10^{15} \,\mathrm{W/cm^2}$ was used. The dashed green vertical and horizontal lines indicate the photoelectron momenta after single ionization of He by absorbing one and two photons.  The dashed black vertical and horizontal lines indicate the photoelectron momenta after ionization of He$^+$ by absorbing one and two photons.}
\label{fig:spdi}
\end{figure}
SPDI is yet another effect where TDDFT does not work in practice, especially if one is interested in correlated photoelectron spectra, for which no density functional is known.

If $\hbar \omega>\abs{E_0^{\mathrm{He}}}$ one photon can fully ionize a helium atom.  However, electron-electron interaction is required in order to share the photon energy absorbed by one electron with another electron. From energy conservation, one obtains
\begin{equation}
 E_\mathrm{kin}^{(1)}+E_\mathrm{kin}^{(2)}=\hbar \omega+E_0^{\mathrm{He}},
\end{equation}
where $E_\mathrm{kin}^{(i)}$ is the kinetic energy of the $i$-th photoelectron. As a consequence, one expects a ring of radius $p=\sqrt{2\iek{\hbar \omega+E_0^{\mathrm{He}}}}$  in correlated photoelectron momentum spectra. If both electrons are emitted in the same direction it is very improbable that one will measure both electrons with the same kinetic energy due to Coulomb repulsion. It is more likely that one electron will have a higher kinetic energy than the other.
% Due to Coulomb repulsion the second electron accelerates the first and slows down.
Thus, we expect the probability along the SPDI ring to vary. In fact, this is seen in Fig.~\ref{fig:spdi}. There is a minimum on the SPDI ring if both photoelectrons have the same energy and are emitted in the same direction.

An atom can simultaneously absorb also two and more photons. If $n$ is the number of  photons which are simultaneously absorbed then the atom can be fully ionized if  $n\hbar \omega>\abs{E_0^{\mathrm{He}}}$. Thus, if the photon energy  $\hbar \omega>\abs{E_0^{\mathrm{He}}}$, rings of radii $p(n)=\sqrt{2\iek{n\hbar \omega+E_0^{\mathrm{He}}}}$ with $n\in\fiek{1,2, 3, ...}$  are expected in correlated photoelectron momentum spectra.  The probability to simultaneously absorb multiple photons  decreases exponentially with the number of photons.  Three rings can be identified in  Fig.~\ref{fig:spdi}, and some traces of a fourth one. In order to observe more rings (within a dynamic range of ten orders of magnitude, as in  Fig.~\ref{fig:spdi}) the laser intensity has to be increased.
\begin{figure}[t]
\centering \includegraphics[width=0.93\columnwidth]{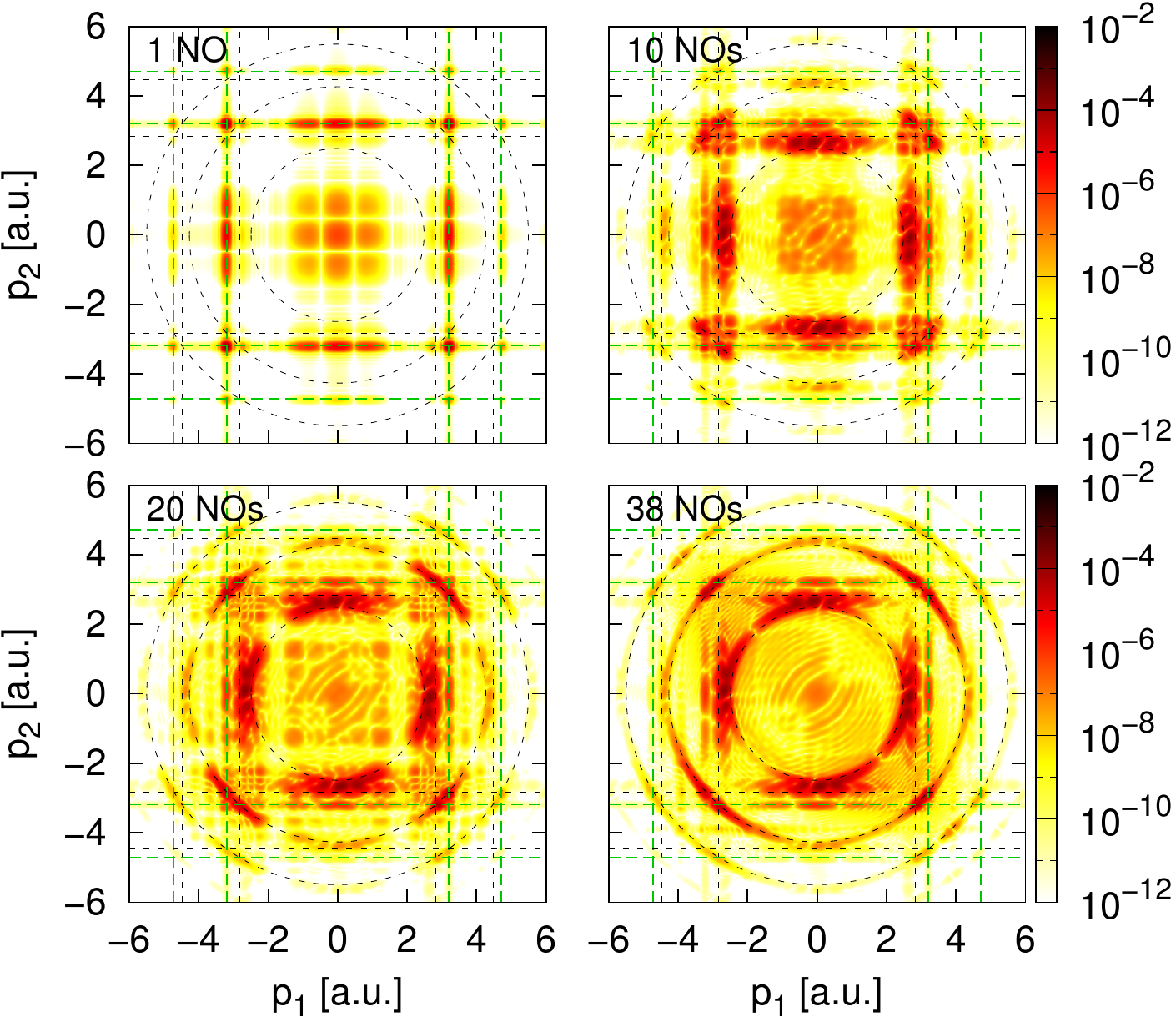}
\caption{(Color online) Correlated photoelectron momentum spectra obtained from the first $i$ spatial NOs for $i=1,10,20,38$, calculated from the TDSE wavefunction at the end of pulse. Same laser pulse as in Fig.~\ref{fig:spdi}, dashed lines having same meaning.}
\label{fig:spdi_no}
\end{figure}

The dashed vertical and horizontal lines in Fig.~\ref{fig:spdi} indicate the expected photoelectron momenta after single ionization of He (green) and He$^+$ (black) by one and two photons.  An enhanced ionization probability is observed when dashed lines of different color cross the higher-order rings ($n=2,3, \ldots$), corresponding to sequential double ionization. The probability is smeared out due to electron-electron interaction, especially if the electrons are emitted in the same direction. The correlated photoelectron momentum spectra were calculated by applying a filter in position-space \cite{florian2, tdrnot3} instead of projecting out all bound and singly ionized states. As this is not a rigorous approach to calculate photoelectron spectra, traces of bound and singly excited states are still visible in Fig.~\ref{fig:spdi}.
\begin{figure}[t]
\centering \includegraphics[width=0.93\columnwidth]{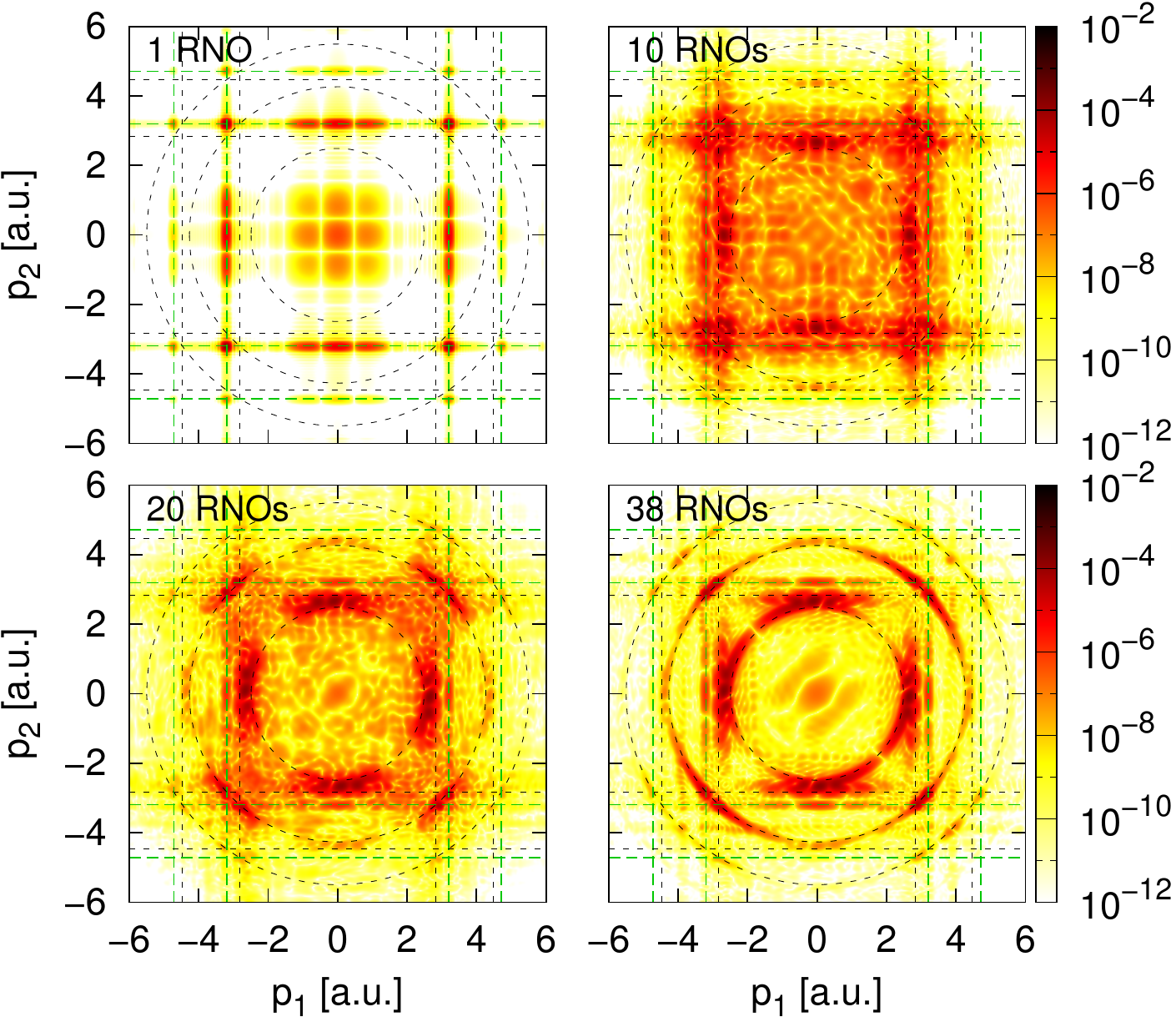}
\caption{(Color online) Correlated photoelectron momentum spectra obtained by TDRNOT/MCTDHF with one to 38 spatial RNOs/determinants. Both methods yield exactly the same spectra when the same number of RNOs/determinants is used. Same laser pulse as in Fig.~\ref{fig:spdi}, dashed lines having same meaning.}
\label{fig:spdi_rno}
\end{figure}

To estimate the minimal number of spatial NOs or determinants required, one may take the first $N_\circ^\mathrm{spat}$ NOs with the highest ONs in \eqref{gamma1inNOs}, calculated from  the exact  TDSE wavefunction at the end of the laser pulse, to evaluate the observable of interest. Note that this corresponds to a hypothetical TDRNOT simulation  without truncation error \cite{tdrnot2,tdrnot3,tdrnot4,tdrnot5}, i.e.,  with an infinite number of NOs taken into account for propagation, but only the dominating $N_\circ^\mathrm{spat}$  NOs used to calculate observables. Comparing  Figs.~\ref{fig:spdi} and \ref{fig:spdi_no} we find that $N_\circ^\mathrm{spat}=38$ NOs are required to accurately  reproduce the SPDI ring $n=1$. $N_\circ^\mathrm{spat} \gg 1$ indicates that SPDI is a very correlated process, and differential, correlated photoelectron momentum spectra are correlation-sensitive observables. It is interesting to investigate how many RNOs for TDRNOT  (or determinants in MCTDHF) are necessary to describe SPDI. In actual TDRNOT/MCTDHF calculations there is a truncation error so that it is expected that more NOs/determinants are needed to reproduce the correlated photoelectron spectra with TDRNOT/MCTDHF than in Fig.~\ref{fig:spdi_no} where the NOs were calculated from the TDSE wavefunction. Such TDRNOT calculations of correlated photoelectron spectra in the context of nonsequential double ionization (NSDI) were pursued in Ref.~\cite{tdrnot3} where twice as many RNOs were found to be necessary for propagation to obtain similar results.
 TDRNOT/MCTDHF results for SPDI are shown in Fig.~\ref{fig:spdi_rno} for selected
   $N_\circ^\mathrm{spat}$  from 1 to 38. At 38, the major features of the
   correlated double-photoelectron spectrum are converged. Quite
   surprisingly, the convergence behavior is only slightly worse than
   that of the TDSE simulation when restricted to the respective $N_\circ^\mathrm{spat}$ in Fig.~\ref{fig:spdi_no}. Thus the truncation error does not play a crucial
   role for SPDI.
This is probably because the laser pulses used for SPDI are of higher frequency and much shorter than in NSDI so that erroneously positioned and unphysical doubly excited states in the two-electron continuum \cite{tdrnot2} due to truncation play a minor part.

\begin{table*}[htbp]
\caption{Time (in seconds) required to calculate correlated photoelectron spectra for SPDI with TDRNOT, MCTDHF, and the TDSE. The calculations were performed on 4 cores of an i5-3570 processor using $N_x = 1024$ grid points in each spatial direction.}
\begin{small}\begin{center}
\begin{tabular}{ccccccccc}
\hline\hline
$ N_\circ^\mathrm{spat}$ &  & \hphantom{11} TDRNOT \hphantom{11} & \hphantom{11}MCTDHF\hphantom{11} & \hphantom{11}TDSE\hphantom{11} & \hphantom{11} MCTDHF/TDRNOT \hphantom{11} & \hphantom{11} TDRNOT/TDSE \hphantom{11} \\ \hline
\multicolumn{ 1}{c}{1 } & total time & 0.82 & 1.29 & 46.99 & \bf{1.6} & \bf{0.02} \\ \cline{ 2- 7}
\multicolumn{ 1}{c}{} & average $\Delta t$ & 0.0094 & 0.0093 & 0.05 & 1.0 & 0.19 \\ \hline
\multicolumn{ 1}{c}{10 } & total time & 9.32 & 247.06 & 46.99 & \bf{26.5} & \bf{0.20} \\ \cline{ 2- 7}
\multicolumn{ 1}{c}{} & average $\Delta t$ & 0.0079 & 0.0102 & 0.05 & 1.3 & 0.16 \\ \hline
\multicolumn{ 1}{c}{20 } & total time & 51.80 & 6251.75 & 46.99 & \bf{120.7} & \bf{1.10} \\ \cline{ 2- 7}
\multicolumn{ 1}{c}{} & average $\Delta t$ & 0.0053 & 0.0080 & 0.05 & 1.5 & 0.11 \\ \hline
\multicolumn{ 1}{c}{\hphantom{1}38 \hphantom{1}} & total time & 467.69 & 190588.00 & 46.99 & \bf{407.5} & \bf{9.95} \\ \cline{ 2- 7}
\multicolumn{ 1}{c}{} & \hphantom{1} average $\Delta t$ \hphantom{1} & 0.0031 & 0.0058 & 0.05 & 1.9 & 0.06 \\ \hline
\end{tabular}
\end{center}\end{small}
 \label{tab:1}
\end{table*}

\subsection{Computational effort}
As already mentioned, any choice of the hermitian operator $\hat g(t)$ will lead to the same results (for a given number of RNOs/determinant) if one technically manages to solve the corresponding EOM.   In practice, the simulations benefit from a gauge choice leading to EOM with good numerical properties. For instance, small matrix elements  $g_{mn}(t)$ usually allow for larger time steps. Thus, by setting $g_{mn}(t)=0$, slightly larger average time steps can be used in MCTDHF than in TDRNOT, as visible from Table~\ref{tab:1}. However, comparing run times one finds that TDRNOT is nevertheless much faster. This is because the analytically known expansion coefficients  $\gamma_{2,ijkl}(t)$ for a two-electron system form a sparse matrix in the NO basis but a dense one in  MCTDHF. Hence, much less matrix elements need to be calculated in TDRNOT where for $2$-electron systems the numerically costly parts of the computations are found to scale as $\tau_{\mathrm{TDRNOT}}\sim N^2_\circ N_t N_x \log(N_x)$ {\em vs} $\tau_{\mathrm{MCTDHF}}\sim N^4_\circ N_t N_x$. Here,  $N_x$ denotes the number of grid points and $N_t$ the number of time steps (different gauges lead to different $\Delta t$ in adaptive propagation schemes). Note that in Refs.~\cite{tdrnot3, tdrnot4} we reported that $\tau_{\mathrm{TDRNOT}}$ contains also a term $\sim N^3_\circ$. 
	However, reduction to $\sim N_\circ^2$ is possible by calculating the sum
	over $p$ in $\mat{B}_{nk}$ prior to the orbital scalar product, as indicated in \eqref{eq:Bnk1}. This comes at no additional cost since the EOM \eqref{eq:eom-rno}
	requires $\sum_k \mat{\hat{C}}_{nk} |\tilde k\rangle$ anyway.

The TDRNOT calculation with 38 RNOs is about 10 times slower than the TDSE. Hence, for the 1D model helium atom TDRNOT does not really offer computational gain. However, as $\tau_{\mathrm{TDSE}}\sim  N_t N_x^2$ TDRNOT becomes superior with increasing $N_x$. Similarly, TDRNOT should be superior for simulations of He in full dimensionality. Unfortunately, it is still unclear if there is any computational gain in TDRNOT over MCTDHF for more than two-electrons. A crucial point here is whether available functionals for $\gamma_{2,ijkl}(t)$ such as the previously mentioned PNOF5e \cite{PNOF5e} and PNOF6($N_c$) \cite{PNOF6}, which are exact in the 2-electron limit, perform well in practice for $N>2$.

\section{Conclusion}
\label{sec:concl}
In this work, we tested further the recently introduced time-dependent renormalized-natural-orbital theory (TDRNOT) on single-photon double ionization (SPDI) of a numerically exactly solvable model helium atom. We showed how TDRNOT is related to multi-configurational time-dependent Hartree-Fock (MCTDHF). We also compared the performance of MCTDHF and TDRNOT, showing that TDRNOT is much faster. Unfortunately, the huge speedup over MCTDHF holds only for two-electron systems. The question whether there is any gain of using TDRNOT over MCTDHF for more-electron systems still needs to be answered and is subject of future work.

\section*{Acknowledgment}
We thank Sven Krönke for inspiring discussions.
This work was supported by the SFB 652 of the German Science Foundation (DFG).

\bibliography{bibliography}

\end{document}